# Liquidity Costs, Idiosyncratic Volatility and Expected Stock Returns

M. Reza Bradrania[a,*], Maurice Peat[b], Stephen Satchell[b,c]
[a]School of Commerce, University of South Australia, Adelaide 5001, Australia
[b] The University of Sydney Business School, Sydney 2006, Australia
[c]Trinity College, Cambridge University, Cambridge CB2 1TQ, UK

*"This is an Accepted Manuscript of an article published by Elsevier in International Review of Financial Analysis in December 2015, available online: https://www.sciencedirect.com/science/article/abs/pii/S1057521915001520 ."*

**Abstract:** This paper considers liquidity as an explanation for the positive association between expected idiosyncratic volatility (IV) and expected stock returns. Liquidity costs may affect the stock returns, through bid-ask bounce and other microstructure-induced noise, which will affect the estimation of IV. We use a novel method (developed by Weaver, 1991) to eliminate microstructure influences from stock closing price-based returns and then estimate IV. We show that there is a premium for IV in value-weighted portfolios, but this premium is less strong after correcting returns for microstructure bias. We further show that this premium is driven by liquidity in the prior month after correcting returns for microstructure noise. The pricing results from equally-weighted portfolios indicate that IV does not predict returns either before or after controlling for liquidity costs. These findings are robust after controlling for common risk factors as well as analysing double-sorted portfolios based on IV and liquidity.

We thank Ben Marshal, Nick Nguyen, Philip Molyneuxand participants at the 11[th]Infiniti Finance conference and Macquarie University research seminars for their insightful comments and suggestions. Special thanks to Terry Walter for his detailed comments.

*Corresponding author.
School of Commerce, University of South Australia
Adelaide 5001, Australia
Email: reza.bradrania@unisa.edu.au
Phone: +61 8 8302 0523
Fax: +61 8 8302 0992.

# 1. Introduction

The pricing ability of idiosyncratic volatility (IV) in asset returns has been at the centre of several studies in the recent asset pricing literature. Portfolio theory suggests that IV is diversified away. The capital asset pricing model (CAPM) predicts that only systematic market risk is priced because all investors hold the market portfolio in equilibrium and so IV is diversified away. However, in reality, investors do not fully diversify their portfolios[1]. There are theories that suggest IV should be priced if investors require compensation for bearing this undiversified risk. For example, Levy (1978) shows that given the assumption of under-diversification IV affects asset prices. Merton (1987) also provides theoretical evidence that there is a positive relation between IV and expected returns when investors do not fully diversify their portfolios.

The empirical results on the existence of the relationship between IV and the cross section of expected stock returns are not consistent either. There are two strands of empirical literature on the pricing of IV. The first strand is based on the *expected volatility* where a long time-series (of monthly return data) is employed to estimate IV (Tinic and West, 1986; Spiegel and Wang, 2005; Malkiel and Xu, 2006; Brockman and Schutte, 2007; Chua et al., 2007; Fu, 2009; Chichernea et al., 2015). These studies have reported a positive relationship between IV and expected returns. The second strand is based on the *realized volatility* estimated using a shorter time-series window (of daily return data) to estimate IV. The studies based on realised volatility have reported a negative (Ang et al., 2006; 2009 and Cotter et al., 2015) or insignificant relationship between IV and expected returns (Bali and Cakici, 2008; Boyer et al., 2010).

Resolving this inconsistency on the role of IV in stock returns is important and has recently generated a rapidly growing literature. Bali and Cakici (2008) provide evidence that shows the *negative* association between expected returns and lagged IV of Ang et al. (2006, 2009) is driven by small firms. Huang et al. (2010) and Han and Lesmond (2011) demonstrate that this negative relationship is driven by return reversals and liquidity. Huang et al. (2010) further show that the *positive* relation between expected IV and returns is robust after controlling for return reversals. Cotter et al. (2015) find that idiosyncratic volatility is significantly negatively priced in stock returns only in down-markets.

---

[1] For several reasons investors may not hold perfectly diversified portfolios. Goetzman and Kumar (2008) show that, based on a sample of more than 62,000 household investors in the period of 1991-1996, more than 25 percent of the investor portfolios contain only one stock, over half of the investor portfolios contain no more than three stocks, and less than 10 percent of the investor portfolios contain more than 10 stocks.

We contribute to this developing literature by providing further evidence to resolve the apparent inconsistency in the association of IV and expected returns. In the spirit of Han and Lesmond (2011) we concentrate on the measurement of IV and show that liquidity costs can explain the positive relationship between expected IV and expected returns reported in several papers.

Spiegel and Wang (2005) investigate the negative correlation between liquidity and IV, and examine whether one variable might be responsible for the premium documented for the other. They show that while there is a premium for each of them, the effect of IV is much stronger and that IV dominates liquidity in explaining the variation of cross-sectional average returns. This result is in line with the work of Constantinides (1986) who shows that the effect of liquidity costs may be confounded with that of risk.

Liquidity explanations for the documented IV premium have been considered and rejected in several studies (Chua et al., 2007; Ang et al., 2006; 2009; Fu, 2009; Huang et al., 2010). In all these studies, a liquidity variable has been included in the cross-sectional tests or has been used in double-sorted portfolios to account for the liquidity effect.

In this paper, we take into account liquidity costs and re-examine the existence and significance of the relationship between expected IV and expected returns. We focus on the estimation of IV before investigating liquidity level as a possible explanation for the volatility premium documented in the literature. Han and Lesmond (2011) theoretically show that the bid-ask microstructure effect on asset returns yields an inflated estimate of IV. We use a new method to eliminate the microstructure effects of bid-ask bounce and other transient errors in closing price-based returns before estimating expected IV[2]. Then we examine directly if liquidity affects the pricing ability of IV in asset pricing tests.

Fisher (1966), Blume and Stambaugh (BS) (1983) and Black (1986) show that observed stock prices can be regarded as the sum of unobservable efficient prices and noise attributable to microstructure effects. BS (1983) extend the work of Fisher (1966) and show that due to Jenson's inequality microstructure noise induces upward bias in observed stock returns. The microstructure-induced noise in returns increases the variation of returns (Asparouhova et al., 2010) which results in an inflated IV estimate. Therefore, ranking stocks and constructing testing portfolios based on biased return estimates might be misleading. Since the microstructure-induced bias is cumulative (Fisher et al., 2010), it can potentially affect portfolio returns and, hence, their relative performance. Moreover, as Asparouhova et al. (2010) point out, the upward bias in returns in the standard regression-based tests in asset pricing induces noisy slope estimates and inflated premia for explanatory variables that are cross-sectionally correlated with the amount of noise in prices. Since there is a non-zero correlation between biased returns and empirical estimates of volatility, we anticipate that the noise associated with microstructure influences might be the reason for the premium reported for IV in the recent literature.

---

[2] As Asparouhova et al. (2010) point out the bias due to microstructure noise is relevant to both monthly and higher frequency (e.g. daily) data because the absolute bias is independent of the return measurement interval.

We use a methodological correction developed by Weaver (1991) and implemented by Fisher et al. (2010) to correct the returns that are used to estimate IV and to compute the average returns of test portfolios. Then, using the corrected returns, we examine if liquidity can explain the premium attributed to IV.

Fisher et al.'s (2010) method applies to equally-weighted portfolios and asymptotically eliminates all random transient errors in portfolio returns. Based on this method, dividing a current two-period average portfolio gross return (one plus the observed return) by prior one-period average portfolio gross return results in an unbiased estimate of the current true one-period gross return. The bias is removed because the average bias in observed returns is attributable to pricing errors at the beginning of the holding period (e.g., BS, 1983). In this method, the returns in the denominator and numerator include the beginning of the period, therefore the bias will cancel out, leaving asymptotically unbiased estimates of true arithmetic returns.

We follow Malkiel and Xu (2006) and estimate individual stock expected IV using the portfolio membership approach suggested by Fama and French (1992) to estimate betas. In particular, we construct equally-weighted rebalanced portfolios, and estimate idiosyncratic volatilities for the portfolios based on the Fama–French three-factor model before and after correcting the underlying returns for microstructure influences. Then, we assign the portfolio volatility estimates to each of the constituent stocks for every month and test the pricing ability of IV using value-weighted and equally-weighted portfolios.

The pricing results of the value-weighted portfolios, before controlling directly for the liquidity level, show that there is a premium for IV before and after correcting returns for microstructure noise. However, this premium is less prevalent, and the positive association between IV and returns is attenuated after correcting for microstructure bias in estimation of

IV. The positive association between expected IV and returns are consistent with several prior studies (Tinic and West, 1986; Malkiel and Xu, 2006; Brockman and Schutte, 2007; Chua et al., 2007; Fu, 2009; Huang et al., 2010) that report a positive premium for IV in value-weighted portfolios.

We further show that this premium is driven by illiquidity in the prior month after correcting returns for microstructure noise. The significant alpha for the zero-investment portfolio with respect to the Carhart (1997) four-factor and Fama-French (1993) three-factors disappears when we include the illiquidity level at the time of portfolio construction in the time series regression. The slope for the illiquidity level is positive and statistically significant at conventional levels, a result which is consistent with the positive premium for illiquidity documented in the literature. This finding holds when we control for the illiquidity level by double-sorting stocks based on illiquidity and IV.

This finding is consistent with that of Han and Lesmond (2011) who use quote midpoint-based returns, which may reflect true returns, instead of closing price-based returns to estimate IV and test its pricing ability. They demonstrate that the negative relation between expected returns and realised IV of Ang et al. (2006, 2009) is driven by liquidity. Their results indicate that although the pricing ability of IV is stronger when IV is estimated using closing returns, it is still significant when quote midpoint returns are used. However, quote midpoint returns are not a good proxy for true returns since liquidity providers set quote prices in a way that moves the midpoints from the true value of the asset in order to be compensated for liquidity costs risks (Ho and Stoll, 1980). We use closing price-based returns, rather than proxies for true returns, to explain positive association between expected IV and returns.

We also examine the pricing ability of IV in equally-weighted portfolios and find that IV cannot predict returns, regardless of the liquidity effect, before or after correcting for microstructure noise. This is consistent with Bali and Cakici (2008) and Huang et al. (2010) who find that IV does not have pricing power on equally-weighted portfolios. However, our correction for bid-ask bounce and other microstructure noise attenuates the magnitude of the statistically insignificant premium for IV in equally-weighted returns.

While our paper falls into the part of the literature that tries to explain the inconsistent results for the IV premium, it is different from studies that provide explanations for a negative association between returns and realised IV. We contribute to the literature in several ways. First, our results confirm that liquidity costs can explain the positive association between expected returns and expected IV documented in the literature for value-weighted portfolios. Second, we employ a novel method to correct closing price-based returns for bid-ask bounce and other microstructure-induced noise which make it possible to utilise closing price-based returns to estimate IV and conduct asset pricing tests. Third, our study provides an alternative way to control for liquidity costs in empirical studies; correcting the returns for bid-ask bounce before estimating any variable measured based on returns. This is particularly important in studies that use IV as a control variable in their empirical tests. Forth, we confirm results of earlier studies that there is no premium for IV in equally-weighted portfolios and that the weighting scheme impacts the pricing ability of IV. Fifth, our finding that there is no premium for IV in equally-weighted portfolios as well as value-weighted portfolios, after controlling for liquidity costs, reconciles earlier studies that report different results in different portfolio settings. Finally, we provide further evidence on the impact of liquidity on expected returns and its pricing ability. One explanation for this finding is that as liquidity has an information component related to future macro fundamentals (Naes et al., 2011) investors consider illiquidity risk when they seek to move away from highly volatile, illiquid stocks. This 'flight to liquidity' impacts asset prices and the premium required by investors. Confirming the pricing power of liquidity that is well established in the literature, our findings have implications in portfolio construction and investment strategies for practitioners and academics.

The paper proceeds as follows. Section 2 elaborates the origins of microstructure noise in closing prices and provides possible solutions to reduce its influence. Section 3 presents the data and explains the construction of the variables. Empirical methods are explained in Section 4, and results are provided and discussed in Section 5. Section 6 offers concluding remarks.

## 2. Microstructure-induced noise in closing prices

Microstructure noise in prices may occur for several reasons. Probably the most important one is bid-ask bounce (BS, 1983). Market buy orders are usually completed at an average price that is higher than the true value of the asset, while market sell orders are completed at an average price that is less than the true asset value. However, the existence of a bid-ask spread does not necessarily imply noise in prices. Glosten and Milgrom (1985) provide a model that shows the spread can arise purely because of asymmetric information. In general, bid-ask bounce is due to the non-information components of spreads, such as processing costs, inventory costs, and potential market-making rents. Huang and Stoll (1997) show that asymmetric information accounts for less than 10 percent of the bid-ask spread in their sample of 20 large stocks during 1992. Non-synchronous trading is another well documented source of microstructure noise, when the last trade prices used to compute returns do not necessarily reflect asset values as at the close of trading. It is also possible for large orders to be closed at prices outside the spread, contributing to the noise in prices. Finally, uninformed orders (Black, 1986) and the discrete pricing grid (Fisher et al., 2010) add noise to observed security prices. All these transient errors move prices away from their efficient values, typically resulting in an upward bias in asset arithmetic returns.

Since the absolute bias is independent of the return measurement interval, the bias due to microstructure noise is relevant to both monthly and higher frequency (e.g. daily) data. However, since the bias relative to true mean returns is larger for high frequency observations the issues related to microstructure noise are particularly important in studies that rely on high frequency data. Asparouhova et al. (2010) focus on monthly stock return data based on CRSP dataset and show that microstructure noise bias increases variation of returns which result in upward bias in liquidity premium.

There are number of potential solutions documented in the literature which reduce or eliminate microstructure-induced noise in computing closing returns, as shown in the following subsections.

### 2.1. Quotation midpoint returns

If we assume that the true, efficient prices are equal to the quote midpoint prices, then the returns computed from these midpoint prices could be an empirical solution. However, Ho and Stoll (1980) show theoretically that liquidity providers set the quote prices in a way that moves the midpoints from the true value of the asset so that they can be compensated for inventory management costs and liquidity risks. Another issue with this method is the availability of data. Quotation data for stocks listed on NYSE and NASDAQ are available from the Trade and Quote (TAQ) database from 1993 onward, and, for NASDAQ stocks, the daily closing quotes are available from 1983 onward (Asparouhova et al., 2010). The available quotation data may not be sufficient to conduct reliable statistical analysis in many asset pricing applications due to lack of power in statistical testing.

### 2.2. Adjusted closing returns

BS (1983) show that upward bias in the observed return for a stock is equal to the variance of the noise. If there is a way to estimate this variance then we can adjust the observed returns by deducting this quantity from each return observation. BS (1983) show that if bid-ask spread is the only source of noise, and that it is constant over time for a particular stock and the stock is traded each period, then this variance would be equal to a quarter of squared spread. However, these assumptions are not empirically valid.

### 2.3. Continuously compounded returns

BS (1983) show that, under reasonable assumptions, continuously compounded returns are free from the upward bias due to microstructure noise. This means that if continuously compounded returns are used in asset pricing tests, microstructure noise bias can be avoided. This method can be used only when we test an asset pricing model that predicts the variation of continuously compounded returns. This solution is not appropriate for testing theories that make predictions regarding holding period returns, such as the discrete-time capital asset pricing model – CAPM.

Ferson and Korajczyk (1995) provide some reasons why continuously compounded returns are not suitable for testing discrete-time asset pricing models. They argue that wealth depends on the returns of investors' portfolios, and continuously compounded portfolio returns are not equal to the portfolio weighted average of the continuously compounded returns of the individual assets. They also show that the average continuously compounded return is less than the mean simple return and the difference between the two increases as the variance of the return increases. Dorfleitner (2003) provides evidence that shows betas estimated from continuously compounded returns are different from betas estimated from simple returns. Fisher (1966) demonstrates that short-horizon continuously compounded returns are downward-biased relative to long holding period returns.

### 2.4. Weighting by lagged stock characteristics

BS (1983) show that buy-and-hold portfolios contain a diversification effect that removes microstructure bias after the first period. Portfolio returns are computed as the weighted average of observed returns, where the weights are the share price at the start of the investment horizon. Since the bias can be approximated by the variance of the previous period's error terms, buy-and-hold portfolios reduce the bias because the weights used after the first period have a negative correlation with subsequent observed returns, offsetting the upward bias. While BS (1983) use observed prior-period share prices as weights, other portfolio weighting methods that depend on the prior-period prices and are negatively correlated with observed returns will also reduce the bias (Asparouhova et al., 2010) and Fisher et al.,2010). For example, one can use any weights based on prior-period market capitalisations or returns and construct value-weighted returns. Since prior-period prices have been used to construct the weights and current-month returns there is a negative correlation between the weights and current returns, leading to less bias in observed returns. The intuition is that if the prior-period observed price is greater (less) than true value, the current-period observed return decreases (increases) on average, while the weight increases (decreases). This reduces the microstructure noise.

BS (1983) point out that this method cannot reduce the bias in the first period of the investment horizon. Moreover, this solution is not appropriate for short-term investment horizons which occur in frequently rebalanced, equally-weighted portfolios. Equal-weighting of returns does not induce negative correlation between the weights (which are constant) and observed returns and, hence, does not reduce bias. BS (1983) show that equally-weighted portfolios will contain the average amount of bias across firms, which can be significant. Based on this intuition, Weaver (1991) develops and Fisher et al. (2010) implement a method for removing bias that allows for frequent rebalancing and applies to equally-weighted portfolios. This method divides a two-period average portfolio price relative (one plus the observed return) ending at time *t*, by a one-period average portfolio price relative that ends at time *t-1* resulting in an unbiased estimate of the true one-period price relative (one plus the return at time *t*). The only assumption is that errors in the prices at time *t*, *t-1*, and *t-2* are independent[3]. The reason that this method removes the bias is that the average bias in observed returns is due to pricing errors at the beginning of the holding period. Because of the law of large numbers the expected bias in observed prices at time *t* is zero, leaving only the bias in observed prices at the beginning of the period. In this method the returns in the denominator and numerator include the beginning of the period; therefore the bias will be cancelled out leaving asymptotically unbiased estimates of true returns. This method asymptotically eliminates biases which are due to any random transient errors and is not limited to bid-ask bounce as in previous studies.

Asparouhova et al. (2010) modify the BS (1983) correction based on the intuition of Weaver (1991) and Fisher et al. (2010) and propose a method for consistent estimation of regression slope coefficients in the presence of microstructure noise in cross-sectional analysis in which researchers use weights on the returns data for each firm. They suggest estimating the cross-sectional regression of observed returns on stock characteristics using the weighted least square (WLS) method rather than OLS, where the weights are equal to the prior-period gross returns.

---

[3] This assumption could be deemed reasonable for the low-frequency data. With High-frequency data bid-ask bounce causes negative dependence.

In this paper, we employ the method developed by Weaver (1991) and implemented by Fisher et al. (2010) to correct the returns that we use to estimate IV and compute average returns of test portfolios.

## 3. Data and variable construction

Daily transaction data (price, returns adjusted for splits and dividends, volume, high/ask, low/bid[4]) are taken from the Centre for Research in Security Prices (CRSP) daily file from January 1$^{st}$, 1958 until December 31$^{st}$, 2008 for all the stocks listed on the NYSE. These data are employed to estimate the monthly liquidity measure. Monthly returns and other data required to compute characteristics for the stocks are downloaded from the CRSP monthly file. Data for the three Fama and French (1993) factors (market, size and value) are downloaded from French's website[5]. We use only NYSE stocks to avoid any possibility of the results being influenced by differences in trading protocols between exchanges.

To be included in the monthly cross-sectional analysis, a stock must satisfy the following selection criteria, suggested by Amihud (2002), Hasbrouck (2009) and Holden (2009):

1- It is an ordinary common share (CRSP codes 10 and 11) traded at the beginning and end of the year.

2- It must have enough trading days, at least two, in a month to estimate the liquidity measure.

3- Its return and market capitalisation data for at least 10 months of each year are available from CRSP.

4- It has monthly data on return and market capitalisation at the start and end of the year.

The first criterion restricts the sample to those stocks usually used in asset pricing studies. The second condition ensures that enough data are used to estimate the liquidity measure. The third condition makes the estimated parameters more reliable, and the fourth criterion provides the data required for portfolio construction. The selection is made based on information in the CRSP event file. In addition to the above criteria, following Chalmers and Kadlec (1998), we eliminate Berkshire Hathaway and Capital Cities because of their unwieldy stock prices. We follow Hasbrouck (2009) and do not remove penny stocks (as in Amihud, 2002) since this may bias our analysis towards liquid stocks. Also, as suggested by Fama and French (1992), we exclude financial firms because they usually have high leverage. While high leverage is normal for these companies, in non-financial firms it is generally a sign of distress. This exclusion improves homogeneity across the stocks.

---

[4] High/ask (low/bid) means the highest (lowest) trade price on a trading day or the closing ask (bid) price on a non-trading day (Holden, 2009).
[5] http://mba.tuck.dartmouth.edu/pages/faculty/ken.french/index.html

This screening process yields on average 3035 stocks. The NYSE introduced a decimal pricing regime to some pilot firms from August 28$^{th}$, 2000, and then completely switched to decimal grids on January 29$^{th}$, 2001. We eliminate from the sample the pilot firms that began to be quoted and traded based on the decimal pricing system between August 28$^{th}$, 2000 and the final switching of NYSE to the decimal regime on January 29$^{th}$, 2001. Since estimation of the liquidity measure is based on tick sizes, this filter makes the computation of the liquidity measure consistent across the stocks in each month. This filtering removes 88 pilot firms from the whole sample.

In order to prepare data for the portfolio constructions and liquidity pricing analysis, we construct the following major variables discussed below.

### 3.1. Liquidity measure

We use effective tick4, henceforth EFFT, developed by Holden (2009) to measure liquidity. EFFT is a daily proxy for the effective spread which picks up on two attributes of the daily data: price clustering on trading days and reported quoted spreads for no-trade days. The proxy has two components corresponding to each of these attributes. The first component, effective tick, which is based on observable price clustering, is a proxy for the effective spread. The second component is the average quoted spread from any no-trade days that exist. This augments effective tick by incorporating the information related to no-trade days. First, we review the effective tick and then define the EFFT estimator. Effective tick is based on the idea that the effective spread on a particular day equals the increment of the price cluster on that particular day. For example on a $1/8 fractional price grid, if the spread is $1/4, Holden's model assumes that prices end on even-eights, or quarters. Thus, if odd-eight transaction prices are observed, one must infer that the spread must be $1/8. This implies that the simple frequency with which closing prices occur in particular price clusters (in a time interval) can be used to estimate the corresponding spread probabilities required to infer the effective spread for that interval. For example on a $1/8 fractional price grid, the frequency with which trades occur in four, mutually exclusive price cluster sets (odd $1/8s, odd $1/4s, odd $1/2s, and whole dollars), can be used to estimate the probability of a $1/8 spread, $1/4 spread, $1/2 spread, and a $1

spread, respectively. There are similar clusters of special prices on a decimal price grid (off pennies, off nickels, off dimes, off quarters, and whole dollars) that can be used to estimate the probability of a penny spread, nickel spread, dime spread, quarter spread and whole dollar spread, respectively. In order to construct the effective tick proxy for a time interval, the first step is to compute the frequency of each price cluster within that time interval. Take $S_t$ as the realisation of the effective spread at the closing trade of day $t$ and assume that $S_t$ is randomly drawn from a set of possible spreads $S_j$ (for example in \$1/8 fractional price grid, $S_1$ = \$1/8 spread, $S_2$ = \$1/4 spread, $S_3$ = \$1/2 spread and $S_4$ = \$1 spread) with corresponding probabilities $\gamma_j$, where j=1,2,…,J and $S_1 < S_2 < … < S_J$. Let $N_j$ be the observed number of trades on odd tick prices corresponding to the jth spread using only positive-volume days in the time interval. The probability of trade prices ($F_j$), corresponding to the jth spread is estimated as:

$$F_j = \frac{N_j}{\sum_{j=1}^{J} N_j} \qquad j=1,2,...,J \qquad (1)$$

Since $F_j$ is based on odd ticks, we calculate $U_j$, the unconstrained probability of the jth spread, to take into account trade prices on even ticks as follows.

$$U_j = \begin{cases} 2F_j & j = 1, \\ 2F_j - F_{j-1} & j = 2,3,...,J-1, \\ F_j - F_{j-1} & j = J. \end{cases} \qquad (2)$$

It follows that for all observed unconstrained probabilities $\sum_{j=1}^{J} U_j = 1$, and if $N_j \leq N_{j+1}$ then $U_j$ are well-defined probabilities (non negative values between 0 and 1). However, in small samples this need not be the case and so we define the constrained probability[6] of the *j*th spread ($\gamma_j$) as

$$\hat{\gamma}_j = \begin{cases} Min\left[Max\{U_j, 0\}, 1\right], & j = 1, \\ Min\left[Max\{U_j, 0\}, 1 - \sum_{k=1}^{j-1} \hat{\gamma}_k\right], & j = 2, 3, \ldots, J. \end{cases} \quad (3)$$

---

[6] This probability assumes a higher frequency on higher rounded increments which is true in large sample. However, in small samples reverse price clustering may be realised that causes the unconstrained probability of one or more effective spread sizes to go above 1 or below zero. Thus, constraints are added to generate proper probabilities.

Then, the effective tick proxy is calculated as the probability-weighted average of each effective spread size divided by the average price ($\bar{p}_i$) in time interval $i$:

$$EffectiveTick_i = \frac{\sum_{j=1}^{J} \hat{\gamma}_j S_j}{\bar{p}_i} \quad (4)$$

Holden (2009) incorporates the average of quoted spreads on no-trade days into the effective tick estimator and concludes that the EFFT for the time interval $i$ is the probability weighted average of the effective estimator and the average of the quoted spreads from no-trade days:

$$EFFT_i = \hat{\mu} \times (EffectiveTick_i) + \frac{\begin{cases} (1-\hat{\mu})\frac{1}{NTD}\sum_{t=1}^{NTD} NQS_t & \text{when } NTD \rangle 0, \\ 0 & \text{when } NTD = 0 \end{cases}}{\bar{P}_i} \quad (5)$$

where $NQS_t$ is the quoted spread computed by using reported bid and ask quoted prices on no-trade days, and $\hat{\mu}$ is the estimated probability of a trading day given by

$$\hat{\mu} = \frac{TD}{TD + NTD} \quad (6)$$

where TD and NTD are, respectively, the number of trading days and no-trade days over the time interval, The method to compute EFFT in a decimal pricing system is slightly different from the above which is suitable for fractional prancing grids. In appendix A, we explain how to calculate EFFT under the decimal pricing regime.

In order to compute the liquidity measure we calculate EFFT at the end of the month using daily trade data. According to the price regime of NYSE over the sample, the approach to computing EFFT is slightly different. From June 1957 to January 2001, during which NYSE was using fractional price grid, price increments as small as $1/64 are used. Following Holden's (2009) suggestion for the application of EFFT on a fractional price grid, we assume that there are seven possible daily bid-ask spreads ($1/64, $1/32, $1/16, $1/8, $1/4, $1/2 and

$1). Therefore, there are seven mutually exclusive price cluster sets (odd sixty-fourths, odd thirty-seconds, odd sixteenths, odd eighths, odd quarters, odd halves and whole dollars) corresponding to each spreads.

From February 2001 to December 2008, during which NYSE used a decimal pricing system, tick sizes are $0.01, $0.05, $0.10, $0.25 and $1. So the possible spreads are $0.01, $0.05, $0.10, $0.25 and $1. The corresponding price cluster sets are off-pennies, off- nickels, off-dimes, off quarters, and whole dollars[7]. For each stock-month, we calculate the frequency of each price cluster, the number of no-trade days, the average trade price from trading days, and the average of the quoted spreads in no trade days. Then, we follow equations 1 to 6 for the stocks under the fractional price system to compute the monthly EFFTs. For those under the decimal regime, we follow the same equations with one exception. Instead of using equation 2 to compute the unconstrained probabilities of the effective spreads, we employ equation A-1 in appendix A.

**3.2. IV measure**

IV is the main variable that we employ to examine the liquidity explanation for the expected IV premium. IV for a period is defined as the standard deviation of the innovation of returns beyond what investors expected, given that period's market returns (Spiegel and Wang, 2005). However, theory does not suggest how the market generates its expectation with regard to the variance of the innovation, so that any measure will be model-dependent.

In most empirical papers, it is assumed that the market follows the Fama–French three-factor model. Researchers estimate the standard deviation of the regression residuals based on the three-factor model using either previous-month daily data (Ang et al., 2006, 2009 and Hang and Lesmond, 2011) to estimate *realised* IV, or the previous 36 or 60 monthly observations (Spiegel and Wang, 2005; Malkiel and Xu, 2006) to estimate *expected* IV[8].

The major difference between these two IV measures is that for the value-weighted portfolios there is a negative and significant relation between realised IV and the cross-section of expected returns (Ang et al., 2006; 2009), whereas there is a positive and significant (or insignificant) relation between expected IV and expected stock returns (Spiegel and Wang, 2005; Malkiel and Xu, 2006). We adopt monthly approach and estimate expected IV as a measure of idiosyncratic volatility for the purpose of this study.

---

[7] Off-pennies are penny price points that are not nickels, dimes, or any higher clusters, namely where the last digit of the price is 1, 2, 3, 4, 6, 7, 8, or 9. Off-nickels are nickel price points that are not dimes, quarters, or any higher clusters, namely where the last two digits of the price are 05, 15, 35, 45, 55, 65, 85, or 95. And so on (Holden, 2009).
[8] A wide variety of statistical tests provide strong evidence that the monthly based measure for IV, compared to the daily-based one, is a more accurate proxy for expected future volatility (Bali and Cakici, 2008).

In order to estimate the volatility for individual stocks, we apply the portfolio membership approach used by Malkiel and Xu (2006) to estimate stock volatilities. We first estimate the idiosyncratic volatilities for the monthly portfolios, constructed at the end of June of each year. The portfolio idiosyncratic volatilities are then assigned to the individual stocks included in the corresponding portfolio.

The reasons that we use the portfolio membership method are threefold. First, analysis based on the idiosyncratic volatilities estimated from individual stocks is more prone to the errors-in-variables problem in the same way as betas estimated from individual stocks. Fama and French (1992) use the portfolio membership method to reduce this error in their estimated betas. Malkiel and Xu (2006) employ this method to reduce the error-in-variable problem in their estimated idiosyncratic volatilities for individual stocks. The second reason is that we are going to use a method to correct the estimation for microstructure noise which is based on portfolios rather than individual stocks. Third, as Malkiel and Xu (2006) assert, it is the *undiversified* IV that matters in explaining asset prices. So it makes sense to estimate the idiosyncratic volatilities of undiversified portfolios to test the pricing ability of IV.

At the end of June for each year, stocks are first sorted into quintile portfolios according to their market capitalisation. Then within each size group, stocks are ranked based on their beta estimated over the previous three to five years using the CAPM model, and another five portfolios are constructed. The market return is the CRSP equally-weighted AMEX/NYSE/NASDAQ index, and the risk-free rate is the one-month Treasury bill rate. The 25 portfolios are rebalanced at the end of June of each year based on the information available in June.

We form portfolios based on size and beta because of the high correlation between size and beta (Fama and French, 1992), and size and IV (Malkiel and Xu, 1997 and 2006). Monthly portfolio returns, size and beta are the cross-sectional average of component stock returns, size and beta. The descriptive statistics for these portfolios over the period of 1958 to 2008 are reported in Table 1.

In each beta group, portfolio returns decrease with portfolio size, as we expect. Within each size group portfolio betas decrease, suggesting that small stocks are riskier. Almost half of the portfolios have a beta less than 1, so the average of the portfolio betas is close to 1. On average, there are 33 stocks in each portfolio, which is enough to mitigate the error-in-variable problem in estimating IV, while leaving the portfolios sufficiently undiversified. With similar data,

Campbell et al. (2001, p.25) examine the standard deviation of portfolio returns and find that almost 50 randomly selected stocks are needed to achieve a well diversified portfolio.

In order to estimate idiosyncratic volatilities at current month for the portfolios we follow Spiegel and Wang (2005) and Malkiel and Xu (2006), and use the previous three to five years of monthly data to compute the standard deviation of the innovations in returns from the Fama–French three-factor model[9].

**[Insert Table 1 about here]**

In each month T, the following model is estimated using the previous three to five years' monthly data.

$$r_{it} = \alpha_0 + \beta_i^m (RM-RF)_t + \beta_i^{smb} SMB_t + \beta_i^{hml} HML_t + \varepsilon_{it}, \; t=[T-1,T-60]^{10} \quad (7)$$

where $r_{it}$ is the equally-weighted monthly excess return on portfolio $i$. $(RM-RF)_t$, $SMB_t$ and $HML_t$ are the Fama–French market, size and book-to-market factors, respectively, in month $t$. $\beta_i^m$, $\beta_i^{smb}$ and $\beta_i^{hml}$ are factor loadings corresponding to Fama–French three factors, and $\varepsilon_{it}$ is the error term.

The IV for the current month ($IV_{it}$) is computed as

$$IV_{it} = \sqrt{\frac{\sum_{j=1}^{T} \hat{\varepsilon}_{it-j}^2}{T-k}} \quad (8)$$

where $T$ is the number of observations over the time horizon and $k$ is the number of estimated parameters (four in this case). A stock is included in the sample if at least 36 out of the 60 previous observations are available for estimation. Each month we assign each portfolio's IV to the stocks within the portfolio in that month.

---

[9] In this paper, we do not investigate the relation between expected stock returns and expected idiosyncratic volatilities conditioned on past information or firm-specific variables (e.g. Fu, 2005).
[10] We use at least 36 months data if 60 months data are not available.

Figure 1 shows the time series of the cross-sectional averages of monthly illiquidity (EFFT) and IV. Both IV and illiquidity increase during periods of financial turbulence and economic crises, such as the stock market crash of October 1987. Moreover, the figure shows clearly a high correlation between these two variables.

**[Insert Figure 1 about here]**

## 4. Empirical methods

This section explains the empirical methods used to investigate the premium for IV and a possible liquidity explanation for this premium. As mentioned in the introduction, the estimation of IV may be influenced by microstructure noise in closing returns. In this section we elaborate the method suggested by Fisher et al. (2010) to eliminate this bias and also explain the general approach and pricing methods used.

### 4.1. True returns calculation

We define the concept of true returns as the returns purified of microstructure effects. We use the method suggested by Fisher et al. (2010) to compute true returns for the portfolios used to estimate IV and the portfolios we use to test the pricing ability of IV.

Based on the analysis of BS (1983), Weaver (1991) develops and Fisher et al. (2010) implement a method for removing bias that allows for frequent rebalancing and applies to equally-weighted portfolios. This method divides a two-period average portfolio price relative (one plus the observed return) ending at time $t$, by a one-period average portfolio price relative that ends at time $t$-1 resulting in an unbiased estimate of the true one-period price relative (one plus the return at time $t$). The only assumption in this method is that errors in the prices at time $t$, $t$-1, and $t$-2 are independent.

The independence assumption allows the replacement of expectations of ratios by ratios of expectations so it is correct to think of expectations here as probability limits as the number of assets in the portfolio increases. In keeping with this literature, we continue to refer to inconsistencies as biases and probability limits as expectations.

The reason that this method removes the bias is that the average bias in observed returns is attributable to pricing errors at the beginning of the holding period. Because of the law of large numbers the expected bias in observed prices at time $t$ will converge to zero, leaving only the bias in observed prices at the beginning of the period. In this method returns in the denominator and numerator include the beginning of the period; therefore the bias will cancel out, leaving asymptotically unbiased estimates of true returns.

$$E\left[\frac{1+{}_2\widehat{R}_t}{1+\widehat{R}_{t-1}}\right] = \frac{[1+E({}_2R_t)] \times \overline{[1+\sigma^2(e_{i,t-2})]}}{[1+E(R_{t-1})] \times \overline{[1+\sigma^2(e_{i,t-2})]}} = \frac{1+E({}_2R_t)}{1+E(R_{t-1})} = 1+E(R_t) \qquad (9)$$

where ${}_2R_t$ is a two-period holding return on an equally-weighted portfolio ending at time $t$, $R_{t-1}$ is a one-period return on an equally-weighted portfolio ending at time $t$-1, and $e_{i,t-2}$ is the bias in the observed return at time $t$-2. $\overline{\sigma^2}(e_{i,t-2})$ is the average of the variance of the bias in the observed return at time $t$-2 in an equally weighted portfolio. The variables with carets are observed returns while the ones without carets are true, unbiased, returns.

The true returns of the equally-weighted portfolio can be estimated via

$$1+E(R_t) = \frac{1+{}_2\widehat{R}_t}{1+\widehat{R}_{t-1}} \qquad (10)$$

Since the equally-weighted portfolio has N stocks then

$$1+E(R_t) = \frac{\sum_{i=1}^{N} 1+{}_2\widehat{R}_{it}}{\sum_{i=1}^{N} 1+\widehat{R}_{it-1}} \qquad (11)$$

The two-period returns can be estimated in the nominator using the constituent one-period returns:

$$1+E(R_t) = \frac{\sum_{i=1}^{N}(1+\widehat{R}_{it-1}) \times (1+\widehat{R}_{it})}{\sum_{i=1}^{N} 1+\widehat{R}_{it-1}} \qquad (12)$$

This method eliminates the biases caused by not only bid-ask bounce, as in previous studies, but also any random transient errors in equally-weighted returns asymptotically.

In particular, for each equally-weighted portfolio in month *t*, we use the first lag and second lag of observed returns of the stocks within portfolios and equation 12 to calculate the true returns for the portfolios. These returns are used to estimate stocks' IV corrected for microstructure-induced bias. The procedure to estimate the corrected idiosyncratic volatilities is the same as the method used to estimate biased IV explained in section 3.2, using true returns rather than CRSP closing returns.

### 4.2. Asset pricing tests

We examine whether IV has explanatory power after correcting for microstructure-induced noise and controlling for the liquidity effect. We employ quintile equally-weighted and value-weighted portfolios and conduct portfolio time-series analysis common in the asset pricing literature. At the end of each month we sort stocks based on their current month's forecast IV before and after correcting for microstructure-induced noise and construct five equally-weighted and value-weighted portfolios. We then form arbitrage portfolios whose returns are the differences between the average returns of the portfolios with the highest and lowest IV. We keep the portfolios for one month and rebalance at the end of the next month. Then we compute portfolio average returns before and after correcting for microstructure-induced noise and compare the results. We also estimate portfolio alphas, as the proxy for risk-adjusted premiums, based on the Fama–French three-factor model and the Carhart (1997) four factor model. If there is any premium for IV we use a simple OLS regression model to examine if this premium can be explained by the variation in liquidity.

## 5. Results and Discussions

In this section, we examine the results of tests to demonstrate whether IV has explanatory power after correcting for microstructure-induced noise and controlling for the liquidity effect. First, we investigate the premium for IV in equally-weighted portfolio returns before and after correction for microstructure noise-induced bias. Next, we conduct pricing analysis on value-weighted portfolios before and after correction for microstructure-induced bias and test whether liquidity can explain the premium detected for IV. In the construction of value-weighted portfolios, current market capitalisation of stocks is used. As discussed in section 2.4, Asparouhova et al. (2010) and Fisher et al. (2010) suggest that portfolio weighting based on the prior-period market capitalisation reduces microstructure noise in returns. We also use lagged market capitalisation in portfolio construction and investigate the IV premium.

## 5.1. Equally-weighted portfolio results

In this section, we construct equally-weighted monthly portfolios based on IV to investigate whether IV is priced. Stocks are sorted into quintiles based on their monthly estimated idiosyncratic volatilities using CRSP closing returns, and equally-weighted portfolios are constructed. Note that the estimated IV and the portfolio average returns are potentially biased because we have not yet corrected the underlying returns for microstructure influences.

At the end of each month, we rank stocks into quintiles based on their current month's forecast IVs estimated using data from the previous five years and form five equally-weighted portfolios. We also construct an arbitrage portfolio with performance equal to the difference between the performance of the highest and lowest IV portfolios. The performance of the arbitrage portfolio is the abnormal return earned by taking a long position in the high IV portfolio and a short position in the low IV portfolio[11]. The portfolios are held for one month and then rebalanced. For each portfolio, we estimate a time-series regression with respect to the Fama–French three-factor model and also the Carhart (1997) four-factor model and report the portfolio Fama–French and Carhart alphas. Table 2 reports the results.

The average returns do not follow any pattern from the highest IV to the lowest IV portfolios, suggesting that there is no relationship between the IV of the equally-weighted portfolios and future returns.

However, after controlling for the Fama–French risk factors, alphas decrease from the lowest IV (0.042 and *t*-statistic=0.45) to the highest IV (-0.351 and *t*-statistic= -3.22). The Fama–French alpha of the arbitrage portfolio, which is long in the highest IV stocks and short in the lowest IV stocks, is -0.87 and insignificant (*t*-statistic = -1.47). The negative sign for the premium is consistent with the finding of Huang et al. (2010) that the equally-weighted return of the portfolio with the lowest IV is higher than that of the portfolio with the highest IV.

**[Insert Table 2 about here]**

When we control for momentum along with the Fama–French factors, the decreasing pattern disappears. The Carhart alpha of the arbitrage portfolio (-0.683) is also less significant (*t*-

---

[11] Whilst this procedure is common in empirical asset pricing, we recognise that not all stocks can be shorted and the cost of shorting is not constant.

statistic= -1.33), suggesting that including a systematic momentum factor reduces the ability of IV to predict one-month-ahead returns[12].

The results in Table 2 also show that illiquidity as measured by EFFT increases monotonically from the low IV portfolio to the high IV portfolio. The equally-weighted spread is 0.484 percent for the low IV quintile and 1.797 percent for the high IV quintile. Given the bid-ask spread effect in returns, this high correlation between IV and the spread shows the effect of the return structure on estimated IV.

Now we use idiosyncratic volatilities corrected for microstructure-induced noise to construct the equally-weighted portfolios. We also compute equally-weighted returns for the quintile portfolios using the Fisher et al. (2010) method to avoid microstructure influences. In other words, not only do we adjust the idiosyncratic volatilities for microstructure noise, we also correct the portfolio returns by applying the Fisher et al. (2010) method on the CRSP closing returns for the IV-based portfolios. The results are presented in Table 3.

**[Insert Table 3 about here]**

The average returns and Carhart alphas do not follow an increasing or decreasing pattern, but the Fama–French alphas decrease from the portfolio with the lowest IV (0.049 and $t$-statistic=0.56) to the one with the highest IV (-0.284 and $t$-statistic=-2.99). The performance of the arbitrage portfolio before and after controlling for common risk factors is also insignificant at conventional levels. These results are similar to those we present in Table 2, before adjusting the returns for microstructure noise, suggesting that IV in equally-weighted portfolios is not priced. This is consistent with the findings of Bali and Cakici (2008) and also Huang et al. (2010) that there is no significant relation between IV and equally-weighted portfolio returns, though they do not take into account microstructure noise bias in returns.

Interestingly, the comparison between Tables 2 and 3 shows that correcting the estimated idiosyncratic volatilities for microstructure noise attenuates the magnitude and significance level of the performance of the arbitrage portfolio. While the average return for this portfolio before controlling for risks is -0.077 percent ($t$-statistic=-0.38) in table 2, it is lower by about 7 basis points after the correction in table 3 (-0.014, $t$-statistic=-0.08). Furthermore, Fama-French and Carhart alphas in table 3 are -0.333 (*t*-statistic=-1.34) and -0.124 (*t*-statistic=-1.11), respectively, which are lower and less significant than those have been reported in table 2.

---

[12] This is inconsistent with the finding of Huang et al. (2010) that after controlling for the past month's returns, the negative relation between average return and lagged IV disappears.

### 5.2. Value-weighted portfolio results

We also examine the pricing ability of IV in value-weighted portfolios. At the end of each month we sort stocks into five value-weighted quintile portfolios based on their monthly expected idiosyncratic volatilities using CRSP closing returns over the previous five years. The idiosyncratic volatilities used in the ranking are biased because of microstructure influences. Current market capitalisations are used as the weights to construct these portfolios. The arbitrage portfolio is the difference between the highest IV portfolio and the lowest IV portfolio. The portfolios are held for one month and then rebalanced. Fama–French and Carhart alphas for each of these portfolios are estimated. The results are tabulated in Table 4.

**[Insert Table 4 about here]**

The average returns increases from the portfolio with the lowest IV (1.181 and *t*-statistic=6.29) to the one with the highest IV (1.771 and *t*-statistic=6.09), suggesting that there is a positive relationship between the IV of the value-weighted portfolios and future returns.

Fama–French alphas also have an increasing pattern with increasing IV and the performance of the arbitrage portfolio is marginally significant (0.33 and *t*-statistic=1.98). After incorporating Carhart's momentum factor, alphas still show an increasing pattern, but the Carhart alpha of the arbitrage portfolio (0.444) is more significant (*t*-statistic = 2.87). This is consistent with the findings of Huang et al. (2010) that controlling for the past month's returns increases the magnitude and significance level of the relation between average return and the lagged IV. These results provide further support for the previous literature that document significant positive association between IV and stock expected returns (Spiegel and Wang, 2006; Fu, 2009 and Malkiel and Xu, 2006). Also, it is consistent with the theory proposed in Merton (1987) which suggests that, since investors are not able to totally diversify IV, they will demand a premium for holding stocks with high IV, and thus stocks with higher IV should deliver higher expected returns.

We now use corrected idiosyncratic volatilities to construct the value-weighted portfolios. The portfolio formation method is the same as the one used to construct value-weighted portfolios using biased idiosyncratic volatilities. Since we use current market capitalisations as the weights to construct these portfolios, the portfolio average returns are not corrected for microstructure noise. Without modification, the Fisher et al. (2010) method to correct closing returns for microstructure-induced bias is only applicable to equally-weighted portfolios. This means that we have only partially removed the microstructure biases from the value-weighted portfolio analysis.

Table 5 reports the results. The premium on the zero-cost portfolio is 0.342 percent (*t*-statistic=2.16) per month which is about 25 basis points lower than the 0.59 percent premium reported in Table 4. The performance of the arbitrage portfolio after controlling for Fama-French (1993) three factors and Carhart (1997) four factors are now attenuated and they are 0.198 percent (*t*-statistic=1.91) and 0.324 percent (*t*-statistic=2.13), respectively. This suggests that Carhart and Fama–French alphas have also been declined by 12 to 13 basis points compared to those in table 4, and while the former is still significant at 5% (*t*-statistic=2.13), the latter is insignificant at 5% level (*t*-statistic=1.91). These findings are consistent with our hypothesis that the pricing power of IV is weaker after controlling for bid-ask bounce and other microstructure noise.

The results in table 5 indicate that after controlling for microstructure effects, the portfolio returns do not have any increasing or decreasing pattern with increasing IV, after controlling for the Fama–French factors and the Carhart factors. However, there is still increasing pattern for the average returns before controlling for risks as it increases from 1.179 percent (for the portfolio with the lowest IV) to 1.342 percent (for the one with the highest IV).

**[Insert Table 5 about here]**

In brief, the results for value-weighted portfolios suggest that while there is a positive association between IV and stock returns, it is less strong, both economically and statistically, after controlling for the bid-ask bounce and other microstructure noise. This supports our conjecture that liquidity costs may impact the relationship between IV and returns before estimation of IV and through microstructure noise. The results from table 5 indicate that there is abnormal returns in taking a long position in the high IV portfolio and a short position in the low IV portfolio even after controlling for common risk factors.

The correction for microstructure noise in value-weighted portfolios was only partial, although the quintile portfolios were constructed using corrected idiosyncratic volatilities we did not correct the value-weighted returns of the quintiles. Furthermore, the monotonic increasing spread in Table 5 suggests that the tradable profits from an arbitrage portfolio that is long in stocks with the highest IV and short in stocks with the lowest IV might be driven by liquidity.

In order to investigate whether liquidity level can explain the premium of this trading strategy, we add the *prior* month's illiquidity level to the four-factor time-series regression model as follows.

$$(High - Low)_t = \alpha_0 + \alpha_1(RM - RF)_t + \alpha_2 SMB_t + \alpha_3 HML_t + \alpha_4 MOM_t + \alpha_5 EFFT_{t-1} + \varepsilon_t \quad (13)$$

where *(High – Low)$_t$* is the month *t* return of the value-weighted portfolio which is long to the portfolio with highest IV (High) and short to the portfolio with the lowest IV (Low). The idiosyncratic volatilities used to construct this portfolio are corrected for microstructure noise using the Fisher et al. (2010) method. *EFFT$_{t-1}$* is the liquidity measure of the high-low portfolio estimated in the prior month which is the time of the portfolios construction based on IV. The *(RM-RF)$_t$*, *SMB$_t$* and *HML$_t$* are the Fama–French market, size and book-to-market factors, respectively, *MOM$_t$* is the Carhart (1997) momentum factor. $\alpha_i$ (*i*≠0) represent the portfolio exposure to the four factors, $\alpha_0$ is a common intercept, and $\varepsilon_t$ is the error term.

If high returns on high IV stocks in the holding month are really driven by their illiquidity level, we would expect that an investment strategy based on IV to show strong comovement with the investment strategy based on the previous month's stock illiquidity. In particular, the abnormal return on the arbitrage strategy based on the difference between the highest and the lowest IV in value-weighted portfolios may be explained by the difference in their previous month's illiquidity.

Table 6 presents the results of the time-series regressions of monthly returns on the value-weighted arbitrage portfolio against three or four factors, where the first two specifications do not control for the previous month's illiquidity of the portfolio and the last two specifications control for illiquidity. If the alphas from the regression are significantly different from zero, the investment strategy based on IV can earn abnormal returns, since the risk loadings on the factors are not sufficient to explain the portfolio's returns.

**[Insert Table 6 about here]**

The intercepts in the first two specifications in Table 6 are similar to the performance of the arbitrage portfolio with respect to Fama–French factors and also the Carhart factors reported in Table 5. They show that the arbitrage portfolio based on sorted IVs earns an abnormal returns of 0.198 (*t*-statistic=1.91) and 0.324 (*t*-statistic=2.13) percent per month after controlling for the three- and four-factor models, respectively. The loadings also indicate that the arbitrage portfolio behaves like small value stocks since it loads positively and heavily on SMB and HML. Thus, a sorting strategy based on IV can have significant tradable returns even after controlling for the conventional risk factors.

The last two specifications in Table 6, where we include the previous month's illiquidity level, indicate that the coefficient on EFFT is positive and statistically significant in both models. This suggests that the holding period return on the sorting strategy based on IV in value-weighted returns can be captured by their previous month illiquidity level. Also none of the intercepts are significantly different from zero, after EFFT is incorporated into the regression model. This result suggests that the value-weighted return difference between the highest IV portfolio and the lowest IV portfolio can be explained by the prior month's illiquidity, while controlling for other conventional risk factors.

To explore the liquidity explanation for the premium of the strategy based on IV we further double-sort the stocks, on illiquidity and then on IV. At the beginning of each month, we rank stocks based on their illiquidity level and then within each illiquidity group we sort the stocks based on their corrected IV. We construct 25 value-weighted portfolios and compute the average returns for each portfolio. The main test of our hypothesis is on arbitrage portfolios. We also form five zero-cost portfolios at each illiquidity level. The performance of these portfolios is the difference between the performance of the portfolios with the highest IV and the one with the lowest IV in each liquidity portfolio. Doing so, we control for the liquidity effect and investigate the premium for each liquidity portfolio. Results are tabulated in Table 7.

**[Insert Table 7 about here]**

The premium is insignificant for all the portfolios and does not exhibit any pattern from low to high IV portfolios. The exception is the arbitrage portfolio on high illiquid stocks, for which there is a premium of 0.9 percent (*t*-statistic=2.68). In general the double-sorting evidence indicates that the premium of the value-weighted sorted strategy based on IV is driven by the illiquidity level.

### 5.3. Lag size value-weighted portfolio results

We did not correct the average returns of the value-weighted portfolios (based on the corrected IVs) by Fisher et al.'s (2010) method because this method is designed to remedy microstructure bias in equally-weighted portfolios. Further, we showed that the premium for the value-weighted arbitrage portfolio based on IV can be explained by illiquidity.

Asparouhova et al. (2010) suggest that using the prior month's market capitalisations to construct value-weighted portfolios removes microstructure noise-induced bias in closing returns. In this section we follow this approach and construct lagged size value-weighted returns. This sort of value-weighting is different from the common practice in which current market capitalisation is used to calculate the weights. The results may not be readily comparable with the value-weighted results we reported above, but they provide further insights into the microstructure noise effect on the premium, which is analysed for the sorting investment strategy based on IV.

At the end of each month we sort stocks into five value-weighted quintile portfolios based on their microstructure bias corrected IV. We use the prior month's market capitalisation as the weights to compute the returns of value-weighted portfolios, so the microstructure influence on the weighted portfolio returns is taken into account. The arbitrage portfolio is the difference between the highest IV portfolio and the lowest IV portfolio. The portfolios are held for one month and then rebalanced. Fama–French and Carhart alphas for each portfolio are estimated. Table 8 presents the results. There is no pattern in returns as IV increases, either before or after controlling for the conventional common risk factors, suggesting that there is no relationship between the IV of lagged size-weighted portfolios and future returns. The Fama–French alpha of the arbitrage portfolio is -0.393 and significant (*t*-statistic = -2.99). However, when we include the momentum factor along with the Fama–French risk factors, the Carhart alpha of the arbitrage portfolio (-0.208) is no longer significant (*t*-statistic= -1.22).

**[Insert Table 8 about here]**

These results suggest that IV is not priced when the value-weighted returns are computed based on lagged market capitalisation. This is consistent with the findings of Bali and Cakici (2008) which show that the ability of IV to predict future returns depends on the weighting method and also on using a screen for size.

## 6. Conclusion

There is a growing literature concerning the inconsistent results for the association of IV and stock expected returns. This is an important issue because the existence of this relationship violates the theories and has direct implications for portfolio construction and investment strategies. In this paper, we investigate the relationship between expected IV and expected stock returns, and examines liquidity as the possible explanation for the positive association between IV and stock expected returns. The initial focus is on the estimation of IV. We correct the estimation of the IV for bid-ask bounce and all the other transient errors in stock closing price-based returns. The pricing tests are conducted on value-weighted and equally-weighted quintile portfolios of US common stocks as representative assets.

The analysis of value-weighted portfolio returns shows that the premium for the IV is positive and significant. However, the pricing ability of IV is significantly weaker, both economically and statistically, after correcting for microstructure bias in the estimation of IV. We show that the sorting investment strategy, based on IV, when the value-weighted portfolio is long in stocks with the highest IV and short in the stocks with the lowest IV, earns about 32 basis points (*t*-statistic=2.13) per month after correcting for the microstructure noise effect and also taking into account the Carhart four factors. This is about 12 basis points lower than the performance before the correction of IV for microstructure noise. These results confirm the positive relation between IV and returns found in earlier studies (e.g. Spiegel and Wang, 2005; Malkiel and Xu, 2006 and more recently Chichernea et al., 2015). However, our further analysis shows that the illiquidity level can explain the ability of IV to positively predict one-month-ahead returns in the zero-investment value-weighted portfolio. The significant alpha for this portfolio with respect to common risk factors disappears when we include the illiquidity level available at the time of portfolio construction in the time-series regression. Confirming the positive impact of illiquidity on returns documented in the literature, the slope for the illiquidity level is positive and statistically significant at conventional levels. This finding is robust after controlling for illiquidity level by double-sorting stocks first based on the illiquidity level and then based on the IV.

Our findings are related to but different from Han and Lesmond (2011). They use quote midpoint-based returns to show that *realised* IV in Ang et al. (2006, 2009), who report the negative relation between IV and returns, has a liquidity-induced bias that overstates the relationship with future returns. In this paper, we focus on *expected* IV which is widely accepted as having a positive association with returns. We use closing price-based returns and show directly that this positive association is driven by liquidity costs. We find support for the notion that 'flight to liquidity' phenomenon induced by the anticipated illiquidity is strong enough to generate a positive relation between IV and expected returns.

The analysis of equally-weighted portfolio returns show that the IV premium is insignificant before and after correcting estimated IV for bid-ask bounce and microstructure effects. The performance of the arbitrage portfolio is lower after this adjustment which is consistent with our findings on value-weighted portfolios. The different pricing results in equally-weighted and value-weighted portfolios, before controlling for the liquidity level, are consistent with the findings of Bali and Cakici (2008) and also Huang et al. (2010) who show that the weighting scheme has a material effect on the pricing ability of IV. While they find an insignificant premium for IV in equally-weighted portfolios, they report significant abnormal returns for the zero-investment value-weighted portfolio constructed based on IV. However, in this paper we show that the positive abnormal return reported in the previous literature for the value-weighted portfolios can be attributed to liquidity costs. We provide a rationale for studies that report mixed results for IV, based on different portfolio settings.

**Appendix A: computation of EFFT in the decimal pricing system**

This section presents the general formula for the EFFT, which works on any decimal (or fractional) price grid. It is based on Appendix A in Holden (2009) and also his example on his website[13].

In a fractional price grid, the price increments overlap completely between adjacent spread levels. For example, all wholes are halves, all halves are quarters, all quarters are eighths, all eighths are sixteenths, etc. However, this does not hold in the decimal system. In the decimal price grid under consideration, all dollars are quarters, all dimes are nickels, and all nickels are pennies, but quarters are different. Two quarters are dimes ($0.50, $1.00) and two quarters are not dimes ($0.25, $0.75). The latter two quarters overlap with nickels (two spread layers down), but not dimes (one spread layer down). Holden (2009) suggests a way to identify and track these overlaps in computing EFFT.

Based on his approach for the decimal system, it is assumed that the possible effective spreads (the $s_j$'s) are $0.01, $0.05, $0.10, $0.25, and $1.00 and $J = 5$.

Let $A_j$ be the *total* number of (trade) prices corresponding to the $j$th spread ($j =1, 2,K, J$). For prices, there are 100 pennies, 20 nickels, 10 dimes, 4 quarters, and 1 dollar, so $A_1 =100$, $A_2 = 20$, $A_3 =10$, $A_4 = 4$, and $A_5 =1$.

Holden defines *special* price increments for the $j$th spread as price increments that can be generated by the $j$th spread, but not by any larger spreads. Let $B_j$ be the number of special prices and corresponding to the $j$th spread ($j =1, 2,K, J$). Also let $O_{jk}$ be the number of price increments for the $j$th spread ($j =1, 2,…K, …, J$) that *overlap* the price increments of the $k$th spread and do *not overlap* the price increments of any spreads between the $j$th spread and the $k$th spread. Table A-1 summarises the $A_j$, $B_j$, and $O_{jk}$ variables under the decimal price regime.

**Table A-1: $A_j$, $B_j$ and $O_{jk}$ for a Decimal Price Grid**

| J | Corresponding spread | $A_j$ | $B_j$ | $O_{jk}$ |
|---|---|---|---|---|
| 1 | $0.01 | 100 | 80 | |
| 2 | $0.05 | 20 | 8 | $O_{21}$=20 |
| 3 | $0.10 | 10 | 8 | $O_{31}$=0, $O_{32}$=10 |
| 4 | $0.25 | 4 | 3 | $O_{41}$=0, $O_{42}$=2, $O_{43}$=2 |
| 5 | $1.00 | 1 | 1 | $O_{51}$=0, $O_{52}$=0, $O_{53}$=0, $O_{54}$=1 |

---

[13] www.kelly.iu.edu/cholden

Then, the general formula for the unconstrained probability of the $j$th spread is

$$U_j = \begin{cases} \left(\dfrac{A_1}{B_1}\right)F_1 & j = 1 \\ \left(\dfrac{A_j}{B_j}\right)F_j - \sum_{k=1}^{j-1}\left(\dfrac{O_{jk}}{B_k}\right)F_k & j = 2,3,\ldots,J \end{cases} \quad (A1)$$

The rest of the effective tick computation is the same as the fraction grid case in the body of the text.

Cotter, J., O' Sullivan, N. and Rossi, F. (2015) 'The conditional pricing of systematic and idiosyncratic risk in the UK equity market', *International Review of Financial Analysis*, 37, 184–193.

Dorfleitner, G. (2003) 'Why the return notion matters', *International Journal of Theoretical and Applied Finance*, 6, 73–86.

Fama, E., and French, K. (1992) 'The cross section of expected stock returns', *Journal of Finance,* 47, 427–466.

Fama, E,. and French, K. (1993) 'Common risk factors in the returns on stocks and bonds', *Journal of Financial Economics*, 33, 3–56.

Fama, E. and Macbeth, J. (1973) 'Risk, return, and equilibrium: Empirical tests', *Journal of Political Economy*, 7, 607–636.

Ferson, W., and Korajczyk, R. (1995) 'Do arbitrage pricing models explain the predictability of stock returns?', *Journal of Business,* 68, 309–349.

Fisher, L. (1966) 'Some new stock-market indexes', *Journal of Business*, 39, 191–225.

Fisher, L., Weaver, D., and Webb, G. P. (2010) 'Removing biases in computed returns', *The Review of Quantitative Finance and Accounting*, 35, 137–161.

Fu, F. (2009) 'IV and the cross-section of expected stock returns', *Journal of Financial Economics*, 91, 24–37.

Huang, W., Liu, Q., Rhee, G., and Zhang, L. (2010) 'Return reversals, IV, and expected returns'*, The Review of Financial Studies,* 23 (1), 147–168.

Glosten, L., and Milgrom, P. (1985) 'Bid, ask, and transaction prices in a specialist market with heterogeneously informed traders', *Journal of Financial Economics,* 14, 71–100.

Goetzmann, W. N., and Kumar, A. (2008) 'Equity Portfolio Diversification', *Review of Finance*, 12 (3), 433–463.

Han, Y., and Lesmond, D.A. (2011) 'Liquidity biases and the pricing of cross-sectional IV', *The Review of Financial Studies,* 24 (5), 1590–1629.

Hasbrouck, J. (2009) 'Trading Costs and Returns for U.S. Equities: Estimating Effective Costs from Daily Data', *The Journal of Finance,* 64 (3), 1445–1477.

Hasbrouck, J. (2006) 'Trading costs and returns for US equities: Estimating effective costs from daily data', Working paper, New York University.

Ho, T., and Stoll, H. (1980) 'On dealer markets under competition', *Journal of Finance,* 35, 259–267.

Holden, C. (2009) 'New low-frequency spread measures', *Journal of Financial Markets*, 12, 778–813.
32

**Figure 1: Cross-sectional Averages of IV and Liquidity**

This plot shows the time-series of the cross-sectional averages of the monthly percentage EFFTs (aggregate EFFT), as a measure of liquidity, and idiosyncratic volatility (IV) for NYSE stocks. The data sample is from June 1958 to December 2008. The first 10 years are used to estimate IVs. Selection criteria to filter the stocks are the same as those employed to construct the EFFT measure and explained in the text. For each firm-month, EFFT is estimated from daily data, and IV is computed as the standard deviation of the regression residual from the three Fama–French factors estimated over a rolling 60-month historical window. Cross-sectional averages of IV and liquidity are the means of the EFFTs and idiosyncratic volatilities across the stocks for each month.

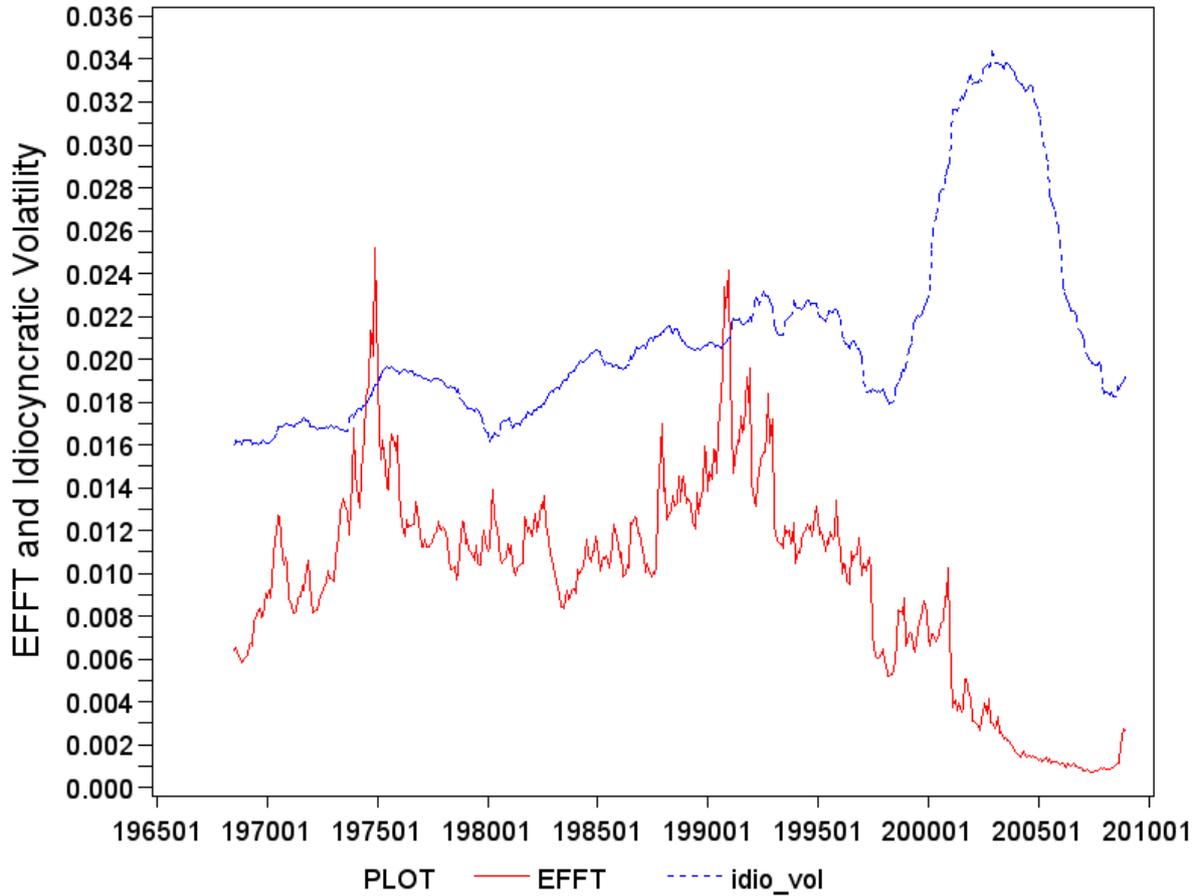

**Table 1: Summary Statistics for 25 Size-beta Equally-weighted Portfolios**

The 25 size-beta equally-weighted portfolios are formed at the end of June of each year by ranking stocks into five equal groups, based on their market capitalisation, and then dividing them into five equal subgroups according to their beta estimated over a rolling 60-month historical window using CAPM. The data sample is from June 1958 to December 2008 and the first five years are used to estimate pre-ranking betas. The table reports the average (over time) of the equally-weighted of the variables for each portfolio. Within each cell, the top number is the average of the monthly equally-weighted portfolio returns in percent. The second number is the average of the portfolio pre-ranking CAPM-betas. The third number is the average of the monthly market value of the portfolio in billions of dollars. The bottom number is the average number of firms within each portfolio.

| Size Group | Returns, [Beta], <Size>, (Number of firms) | | | | | |
|---|---|---|---|---|---|---|
| | Beta Group | | | | | |
| | Lowest | 2 | 3 | 4 | Highest | Mean |
| Small | 1.2 [0.5444] <0.0865> (32) | 1.4 [0.8772] <0.0981> (32) | 1.45 [1.1123] <0.0961> (32) | 1.45 [1.3642] <0.0973> (33) | 1.39 [1.8342] <0.0961> (32) | 1.38 [1.1465] <0.0948> (32) |
| 2 | 1.18 [0.4933] <0.3278> (33) | 1.33 [0.8626] <0.3113> (33) | 1.18 [1.0953] <0.3102> (33) | 1.06 [1.3223] <0.3043> (33) | 1.02 [1.7838] <0.3055> (34) | 1.15 [1.1115] <0.3118> (33) |
| 3 | 1.02 [0.4579] <0.752> (34) | 1.12 [0.7989] <0.7433> (34) | 1.19 [1.0323] <0.7347> (33) | 1.07 [1.2571] <0.7377> (34) | 0.98 [1.7026] <0.7465> (34) | 1.08 [1.0498] <0.7428> (34) |
| 4 | 0.94 [0.4565] <1.8091> (34) | 1.02 [0.7726] <1.8231> (34) | 1.02 [0.994] <1.8078> (34) | 0.95 [1.2113] <1.7782> (34) | 0.84 [1.6504] <1.7572> (34) | 0.95 [1.017] <1.7951> (34) |
| Big | 0.87 [0.4619] <11.2722> (34) | 0.92 [0.7355] <15.454> (34) | 0.98 [0.9121] <11.7979> (34) | 0.85 [1.1008] <11.8495> (34) | 0.68 [1.4706] <10.8434> (34) | 0.86 [0.9362] <12.2434> (34) |
| Mean | 1.04 [0.4828] <2.8495> (33) | 1.16 [0.8094] <3.686> (33) | 1.16 [1.0292] <2.9493> (33) | 1.08 [1.2511] <2.9534> (34) | 0.98 [1.6883] <2.7497> (34) | 1.08 [1.0522] <3.0376> (33) |

**Table 2: Performance of Equally-weighted Portfolios Sorted on Non-corrected IV**

The table reports the attributes of five equally-weighted portfolios constructed based on idiosyncratic volatility (IV) which are not corrected for microstructure-induced noise. Estimation of IV is explained in the text. At the end of each month, stocks are ranked based on their estimated IV and five equally-weighted portfolios are constructed. The portfolios are rebalanced at the end of each month. High-Low is the portfolio which is long to the portfolio with highest IV (High) and short to the portfolio with the lowest IV (Low). The data sample is from June 1958 to December 2008. The first 10 years are used to estimate IVs. Return is the time-series average of the monthly equally-weighted portfolio returns. EFFT and IV are the measures for liquidity and IV, respectively. Fama–French Alpha and Carhart Alpha are the intercepts estimated from the time-series regression of portfolio returns against the Fama–French three factors and also the four factors comprising the three Fama–French factors and the Carhart (1997) momentum factor, respectively. All values are in percentage. *T*-statistics are in parentheses and adjusted for HAC using Newy-West (1987) method with five lags. *, **, *** denote significance at 10%, 5%, and 1%, respectively.

| Rank | IV | EFFT | Returns | Alpha | |
|---|---|---|---|---|---|
| | | | | Fama–French | Carhart |
| Low | 1.457*** | 0.484*** | 0.969*** | 0.042 | 0.086 |
| | (14.1) | (6.25) | (5.37) | (0.45) | (1.37) |
| 2 | 1.742*** | 0.723*** | 1.081*** | -0.035 | -0.009 |
| | (16.55) | (5.38) | (5.26) | (-0.52) | (-0.13) |
| 3 | 2.017*** | 0.832*** | 0.949*** | -0.158 | -0.065 |
| | (16.7) | (5.43) | (3.98) | (-1.6) | (-1.01) |
| 4 | 2.324*** | 0.971*** | 0.931*** | -0.219*** | -0.059 |
| | (14.96) | (7.09) | (3.94) | (-2.68) | (-1.14) |
| High | 3.019*** | 1.797*** | 0.891*** | -0.351*** | -0.12 |
| | (11.44) | (4.78) | (3.1) | (-3.22) | (-1.21) |
| High-low | 1.561*** | 1.313*** | -0.077 | -0.87 | -0.683 |
| | (7.85) | (3.75) | (-0.38) | (-1.47) | (-1.33) |

**Table 3: Performance of Equally-weighted Portfolios Sorted on Corrected IV**

The table reports the attributes of five equally-weighted portfolios constructed based on idiosyncratic volatility (IV) corrected for microstructure noise using the Fisher et al. (2010) method. Estimation of IV and the correction method are explained in the text. At the end of each month, stocks are ranked based on their corrected IV and five equally-weighted portfolios are constructed. The portfolios are rebalanced at the end of each month. High-Low is the portfolio which is long to the portfolio with highest IV (High) and short to the portfolio with the lowest IV (Low). The data sample is from June 1958 to December 2008. The first 10 years are used to estimate IVs. Return is the time-series average of the monthly equally-weighted portfolio returns. EFFT and IV are the measures for liquidity and IV, respectively. Fama–French Alpha and Carhart Alpha are the intercepts estimated from the time-series regression of portfolio returns against the Fama–French three factors and also the four factors comprising the three Fama–French factors and the Carhart (1997) momentum factor, respectively. All values are in percentage. *T*-statistics are in parentheses and adjusted for HAC using Newy-West (1987) method with five lags. *, **, *** denote significance at 10%, 5%, and 1%, respectively.

| Rank | IV | EFFT | Returns | Alpha | |
|---|---|---|---|---|---|
| | | | | Fama–French | Carhart |
| Low | 1.446*** | 0.479*** | 0.991*** | 0.049 | 0.09 |
| | (14.4) | (6.09) | (5.5) | (0.56) | (1.53) |
| 2 | 1.732*** | 0.732*** | 1.106*** | 0.015 | 0.061 |
| | (17.2) | (5.22) | (5.39) | (0.23) | (1.15) |
| 3 | 2.005*** | 0.805*** | 0.999*** | -0.146 | -0.04 |
| | (17.08) | (5.6) | (4.1) | (-1.43) | (-0.59) |
| 4 | 2.312*** | 0.997*** | 0.967*** | -0.174** | 0.002 |
| | (14.82) | (6.9) | (4.04) | (-1.91) | (0.02) |
| High | 2.98*** | 1.793*** | 0.976*** | -0.284*** | -0.033 |
| | (11.86) | (4.87) | (3.45) | (-2.99) | (-0.38) |
| High-low | 1.535*** | 1.314*** | -0.014 | -0.333 | -0.124 |
| | (8.22) | (3.83) | (-0.08) | (-1.34) | (-1.11) |

**Table 4: Performance of Value-weighted Portfolios Sorted on Non-corrected IV**

The table reports the attributes of five value-weighted portfolios constructed based on idiosyncratic volatility (IV), not corrected for microstructure-induced noise. Estimation of IV is explained in the text. At the end of each month, stocks are ranked based on their estimated IV and five value-weighted portfolios are constructed. The portfolios are rebalanced at the end of each month. High-Low is the portfolio which is long to the portfolio with highest IV (High) and short to the portfolio with the lowest IV (Low). The data sample is from June 1958 to December 2008. The first 10 years are used to estimate IVs. Return is the time-series average of the monthly value-weighted portfolio returns. EFFT and IV are the measures for liquidity and IV, respectively. Fama–French Alpha and Carhart Alpha are the intercepts estimated from the time-series regression of portfolio returns against the Fama–French three factors and also the four factors comprising the three Fama–French factors and the Carhart (1997) momentum factor, respectively. All values are in percentage. $T$-statistics are in parentheses and adjusted for HAC using Newy-West (1987) method with five lags. *, **, *** denote significance at 10%, 5%, and 1%, respectively.

| Rank | IV | EFFT | Returns (%) | Alpha | |
|---|---|---|---|---|---|
| | | | | Fama–French | Carhart |
| Low | 1.389*** | 0.353*** | 1.181*** | 0.453*** | 0.429*** |
| | (13.81) | (6.49) | (6.29) | (5.22) | (5.42) |
| 2 | 1.661*** | 0.468*** | 1.389*** | 0.482*** | 0.462*** |
| | (14.14) | (5.23) | (7.27) | (5.98) | (5.81) |
| 3 | 1.95*** | 0.583*** | 1.477*** | 0.519*** | 0.481*** |
| | (15.94) | (5.92) | (6.93) | (5.23) | (5.36) |
| 4 | 2.281*** | 0.604*** | 1.449*** | 0.612*** | 0.525*** |
| | (14.62) | (6.83) | (7.94) | (6.38) | (6.44) |
| High | 2.798*** | 0.99*** | 1.771*** | 0.783*** | 0.873*** |
| | (11.71) | (4.38) | (6.09) | (5.04) | (5.16) |
| High-low | 1.409*** | 0.637*** | 0.59*** | 0.33** | 0.444*** |
| | (7.78) | (4.11) | (2.71) | (1.98) | (2.87) |

**Table 5: Performance of Value-weighted Portfolios Sorted on Corrected IV**

The table reports the attributes of five value-weighted portfolios constructed based on idiosyncratic volatility (IV) corrected for microstructure-induced noise using the Fisher et al. (2010) method. Estimation of IV and the correction method are explained in the text. At the end of each month, stocks are ranked based on their estimated IV and five value-weighted portfolios are constructed. The portfolios are rebalanced at the end of each month. High-Low is the portfolio which is long to the portfolio with highest IV (High) and short to the portfolio with the lowest IV (Low). The data sample is from June 1958 to December 2008. The first 10 years are used to estimate IVs. Return is the time-series average of the monthly value-weighted portfolio returns. EFFT and IV are the measures for liquidity and IV, respectively. Fama–French Alpha and Carhart Alpha are the intercepts estimated from the time-series regression of portfolio returns against the Fama–French three factors and also the four factors comprising the three Fama–French factors and the Carhart (1997) momentum factor, respectively. All values are in percentage. *T*-statistics are in parentheses and adjusted for HAC using Newy-West (1987) method with five lags. *, **, *** denote significance at 10%, 5%, and 1%, respectively.

| Rank | IV | EFFT | Returns (%) | Alpha | |
| --- | --- | --- | --- | --- | --- |
| | | | | Fama–French | Carhart |
| Low | 1.365*** | 0.341*** | 1.179*** | 0.446*** | 0.415*** |
| | (13.73) | (6.14) | (6.83) | (5.82) | (5.88) |
| 2 | 1.646*** | 0.473*** | 1.391*** | 0.544*** | 0.529*** |
| | (16.88) | (5.88) | (7.13) | (7.84) | (6.93) |
| 3 | 1.917*** | 0.555*** | 1.401*** | 0.501*** | 0.436*** |
| | (16.72) | (5.47) | (5.99) | (5.37) | (5.21) |
| 4 | 2.219*** | 0.679*** | 1.430*** | 0.476*** | 0.512*** |
| | (15.13) | (6.93) | (7.32) | (5.71) | (6.08) |
| High | 2.663*** | 0.971*** | 1.521*** | 0.644*** | 0.739*** |
| | (12.75) | (4.21) | (6.75) | (5.19) | (5.73) |
| High-low | 1.1.298*** | 0.63*** | 0.342** | 0.198* | 0.324** |
| | (8.71) | (3.26) | (2.16) | (1.91) | (2.13) |

**Table 6: Test of Illiquidity as the Explanation for the Arbitrage Value-weighted Portfolio based on IV**

The table reports the results of the following time-series regression.

$$(High - Low)_t = \alpha_0 + \alpha_1 (RM - RF)_t + \alpha_2 SMB_t + \alpha_3 HML_t + \alpha_4 MOM + \alpha_5 EFFT_{t-1} + \varepsilon_t$$

where *(High-Low)$_t$* is the month *t* return of the value-weighted portfolio which is long to the portfolio with highest IV (High) and short to the portfolio with the lowest IV (Low). The IVs are corrected for microstructure noise using the Fisher et al. (2010) method. EFFT$_{t-1}$ is the illiquidity measure of the high-low portfolio estimated in the prior month. *(RM-RF)$_t$*, *SMB$_t$* and *HML$_t$* are the Fama–French three factors and *MOM$_t$* is the Carhart (1997) momentum factor, and *ε$_t$* is the error term. The data sample is from June 1958 to December 2008. The first 10 years are used to estimate IVs. *T*-statistics are in parentheses and adjusted for HAC using Newy-West (1987) method with five lags. *, **, *** denote significance at 10%, 5%, and 1%, respectively.

|  | Specifications | | | |
|---|---|---|---|---|
| Variable | 1 | 2 | 3 | 4 |
| Intercept | 0.198* (1.91) | 0.324** (2.13) | 0.098 (0.38) | 0.017 (0.87) |
| RM-RF | 22.841*** (2.98) | 21.14*** (3.22) | 22.429** (2.34) | 20.349*** (2.94) |
| SMB | 72.113*** (8.17) | 72.613*** (8.49) | 71.877*** (7.44) | 71.341*** (8.23) |
| HML | 26.013*** (3.78) | 23.13*** (3.75) | 25.483*** (3.98) | 22.584*** (3.49) |
| MOM |  | -17.871*** (-4.82) |  | -18.127*** (-4.13) |
| EFFT |  |  | 55.732*** (2.98) | 61.837*** (3.11) |

**Table 7: Performance of the Value-weighted Portfolios Double-sorted on Illiquidity and Corrected IV**

The table reports the performance of 25 monthly value-weighted portfolios constructed based on the illiquidity level and the idiosyncratic volatility (IV) corrected for microstructure noise using the Fisher et al. (2010) method. Estimation of IV and the correction method are explained in the text. At the end of each month, stocks are ranked first based on their illiquidity, measured by EFFT, and then within each illiquidity group based on their corrected IV, and 25 value-weighted portfolios are constructed. The portfolios are rebalanced at the end of each month. High-Low is the portfolio which is long to the portfolio with the highest IV (High) and short to the portfolio with the lowest IV (Low) in each illiquidity group. The data sample is from June 1958 to December 2008. The first 10 years are used to estimate IVs. The table reports the time-series average of the monthly value-weighted portfolio returns and in percentage. Fama–French Alpha and Carhart Alpha are the intercepts estimated from the time-series regression of portfolio returns against the Fama–French three factors and also the four factors comprising the three Fama–French three factors and the Carhart (1997) momentum factor, respectively. *T*-statistics are in parentheses and adjusted for HAC using Newy-West (1987) method with five lags. *, **, *** denote significance at 10%, 5%, and 1%, respectively.

|   | Low-IV | 2 | 3 | 4 | High-IV | High-Low | Alpha for High-Low | |
|---|---|---|---|---|---|---|---|---|
|   |   |   |   |   |   |   | Fama–French | Carhart |
| Low-illiquid | 1.141*** | 1.115*** | 1.162*** | 1.298*** | 1.174*** | 0.034 | -0.105 | -0.124 |
|   | (6.16) | (6.63) | (6.1) | (5.38) | (5.54) | (0.27) | (-0.88) | (-1.24) |
| 2 | 1.42*** | 1.356*** | 1.419*** | 1.373*** | 1.309*** | -0.112 | -0.225 | -0.135 |
|   | (6.71) | (6.4) | (6.45) | (7.51) | (5.88) | (-0.79) | (-1.35) | (-0.88) |
| 3 | 1.514*** | 1.748*** | 1.658*** | 1.361*** | 1.536*** | 0.022 | -0.088 | -0.007 |
|   | (7.09) | (7.91) | (6.05) | (6.28) | (5.61) | (0.14) | (-0.65) | (-0.04) |
| 4 | 1.428*** | 1.651*** | 1.654*** | 1.725*** | 1.67*** | 0.242 | 0.077 | 0.084 |
|   | (8.03) | (7.04) | (6.55) | (6.66) | (5.62) | (1.01) | (0.46) | (0.54) |
| High-illiquid | 2.033*** | 2.449*** | 2.236*** | 2.301*** | 2.945*** | 0.912*** | 0.618*** | 0.579** |
|   | (7.7) | (5.43) | (7.32) | (6.11) | (6.14) | (2.68) | (2.58) | (2.02) |

**Table 8: Performance of Lag Size-weighted Portfolios Sorted on Corrected IV**

The table reports the attributes of five lag size value-weighted portfolios constructed based on idiosyncratic volatility (IV) corrected for microstructure noise using the Fisher et al. (2010) method. Estimation of IV and the correction method are explained in the text. At the end of each month, stocks are ranked based on their corrected IV and five value-weighted portfolios are constructed. The weights are based on the market capitalisations of the stocks in the prior month. The portfolios are rebalanced at the end of each month. High-Low is the portfolio which is long to the portfolio with highest IV (High) and short to the portfolio with the lowest IV (Low). The data sample is from June 1958 to December 2008. The first 10 years are used to estimate IVs. Return is the time-series average of the monthly value-weighted portfolio returns and in percentage. EFFT and IV are the measures for liquidity and IV, respectively. Fama–French Alpha and Carhart Alpha are the intercepts estimated from the time-series regression of portfolio returns against the Fama–French three factors and also the four factors comprising the three Fama–French three factors and the Carhart (1997) momentum factor, respectively. *T*-statistics are in parentheses and adjusted for HAC using Newy-West (1987) method with five lags. *, **, *** denote significance at 10%, 5%, and 1%, respectively.

| Rank | IV | EFFT | Returns (%) | Alpha | |
|---|---|---|---|---|---|
| | | | | Fama–French | Carhart |
| Low | 1.389*** | 0.314*** | 0.865*** | 0.069 | 0.036 |
| | (13.98) | (6.56) | (4.76) | (0.99) | (0.63) |
| 2 | 1.721*** | 0.438*** | 0.973*** | 0.027 | 0.039 |
| | (17.07) | (5.63) | (5.11) | (0.33) | (0.38) |
| 3 | 2.008*** | 0.513*** | 0.86*** | -0.083 | -0.097 |
| | (16.34) | (5.32) | (3.67) | (-1.14) | (-1.67) |
| 4 | 2.29*** | 0.603*** | 0.778*** | -0.228** | -0.131 |
| | (14.88) | (6.12) | (3.51) | (-2.52) | (-1.57) |
| High | 2.847*** | 0.942*** | 0.763*** | -0.324*** | -0.171* |
| | (12.45) | (4.96) | (3.53) | (-2.8) | (-1.81) |
| High-low | 1.458*** | 0.628*** | -0.102 | -0.393*** | -0.208 |
| | (8.47) | (3.77) | (-0.57) | (-2.99) | (-1.22) |